\documentclass[twocolumn,prl] {revtex4}

\usepackage{graphicx}
\usepackage{dcolumn}
\usepackage{amsmath}
\usepackage{bm}
\usepackage{lscape}
\begin{document}
\title {Segmental relaxation in semicrystalline polymers: a mean field model for the distribution of relaxation times in confined regimes}
\author{Marco Pieruccini$^{1}$, Tiberio A. Ezquerra$^2$}
\affiliation{$^1$ CNR, Istituto per i Processi Chimico-Fisici, Messina, Salita Sperone Contrada Papardo, sn 98158 Faro Superiore, Messina, Italy}
\affiliation{$^2$ Instituto de Estructura de la Materia, CSIC, Serrano 119, 28006 Madrid, Spain}

%\author{PACS: 64.70.pj, 64.70.qd, 77.22.Gm}
%\affiliation{ }
%\author{Short title: \textbf{Confined segmental relaxation in polymers}}

%\date{\today}
%%%%%%%%%%%%%%%%%%%%%%%%%%%%%%%%%%%%%%%%%%%%%%%%%%%%%%%%%

\begin{abstract}
The effect of confinement in the segmental relaxation of polymers is considered. On the basis of a thermodynamic model we discuss the emerging relevance of the fast degrees of freedom in stimulating the much slower segmental relaxation, as an effect of the constraints at the walls of the amorphous regions. In the case that confinement is due to the presence of crystalline domains, a quasi-poissonian distribution of local constraining conditions is derived as a result of thermodynamic equilibrium. This implies that the average free energy barrier $\Delta F$ for conformational rearrangement is of the same order of the dispersion of the barrier heights, $\delta (\Delta F)$, around $\Delta F$. As an example, we apply the results to the analysis of the $\alpha$-relaxation as observed by dielectric broad band spectroscopy in semicrystalline poly(ethylene terephthalate) cold-crystallized from either an isotropic or an oriented glass. It is found that in the latter case the regions of cooperative rearrangement are significantly larger than in the former.

%KEYWORDS: nano-confinement, $\alpha$ relaxation, polymer crystallization, dielectric response
\end{abstract}

\maketitle

\section{Introduction}

Segmental relaxation is a process whereby a certain initial chain conformation, produced by a thermodynamic fluctuation in an
amorphous region of a polymer, evolves towards a more probable state. This mechanism has a central role in several processes, such as elastic recovery~\cite{ferry}, dielectric response~\cite{kremer}, diffusion of small molecules~\cite{baer},
glass transition~\cite{donth} and local chain readjustments at the boundaries of growing crystalline domains~\cite{strobl}, to mention a few.

Recently, the problem of understanding how confinement affects conformational relaxation has attracted growing
interest~\cite{confinement_1}. In this respect, semicrystalline polymers represent a natural ground where some progress may be
done in this field, since neither melt- nor cold-crystallization lead to complete crystalline states in these materials. At the end of the crystallization process, the remaining amorphous regions cannot convert to crystalline because of the conformational constraints induced by the confining crystal domains through pinning and chain connectivity. In particular, the dynamics of the
$\alpha$-relaxation associated to the segmental motions above the glass transition temperature $T_g$ is strongly affected by
the progressive development of the crystalline phase~\cite{Willi1,Boyd1,tib1}.

From the thermodynamic point of view, the establishment of constraints in the amorphous regions reduces the conformational entropy, causing an increase of the chemical potential $\mu_a$ with respect to the value $\mu_{a0}$ of the unconstrained
liquid. On the other hand, crystallization from an undercooled liquid (either constrained or not) is associated to a decrease of the chemical potential down to a value $\mu_c < \mu_{a0}$.

The crystals reduce chain mobility; this is not only due to the decrease of the volume accessible to the amorphous chains, but also because many of these chains are anchored to the crystals. Thus, crystal growth is always accompanied by an increase of $\mu_a$. In the initial stages of the crystallization process, the difference $\mu_a - \mu_{a0}$ is small compared to the chemical potential drop $\Delta\mu_{ac}\equiv\mu_c - \mu_{a0}$ associated to the amorphous-to-crystalline transition of a monomer. However, as crystallinity increases, the change of $\mu_a - \mu_{a0}$ accompanying further transition of monomers to the crystalline state is so large that it cannot be compensated by the corresponding free energy decrease associated to $|\Delta\mu_{ac}|$. The whole system thus tends to a stationary state beyond which the phase transition cannot proceed further.

On the basis of this idea, the vitrification process observed in the low-temperature crystallization of uniaxially oriented poly(ethylene terephthalate) (PET) has been recently discussed~\cite{PRB}.

Experiments show that the glass transition temperature $T_g$ increases with respect to a completely amorphous material, as an
effect of a finite crystallinity~\cite{Schick,dobbertin}. However, it is important to point out that this variation is always
accompanied by an increase of the temperature interval over which the amorphous component of the system de-vitrifies upon heating. These effects, which can be easily observed by means of differential scanning calorimetry, are particularly evident when a dynamical mechanical thermal analysis in the $\alpha$-relaxation region is performed (see e.g. ref.~\cite{PRB} for the case of PET). This suggests that the emergence of constraints to chain readjustments and the development of inhomogeneities in their character could be inherently connected.

The analysis of the crystallization process by means of broad band dielectric spectroscopy (BDS) gives similar indications.
Recent investigations on the crystallization of poly(propylene succinate) and poly(propylene adipate) (PPS and PPA
respectively)~\cite{Tiberio_PRL,Tiberio_PPS-PPA} represent particularly clear examples of the broadening effect associated to
the increase of constrainings. These studies pointed out that the dominant characteristics of the $\alpha$ relaxation change during crystallization. Namely, the rather fast and comparatively narrow dielectric $\alpha$-process observable in completely amorphous samples is progressively substituted by a much slower and broader process (in frequency) eventually overwhelming the former. In some polymers like PPA and PET~\cite{Tiberio_PPS-PPA,Tiberio_PET,Fukao}, the two relaxations can be observed as distinct, separable processes, one increasing in intensity at the expenses of the other as crystallization proceeds. In other polymers, like PPS~\cite{Tiberio_PRL,Tiberio_PPS-PPA} and poly(L lactide acid)~\cite{Bras,Mantana,Jovan} one observes just one process whose average characteristic relaxation time decreases progressively while
broadening in frequency.

With the aim of describing how the establishment of structural constraints hinders conformational rearrangement and gives rise to time heterogeneity, we shall resort to the concept of cooperatively rearranging region (CRR) originally introduced by
Adam and Gibbs for the description of liquids close to their glass transition~\cite{A_G}.

A CRR is the smallest amorphous domain where a conformational rearrangement may occur without causing any structural change at its boundary. Last condition is in fact a constraint which determines the existence of a free energy barrier $\Delta F$ that a thermodynamic fluctuation needs to overcome to produce a conformational rearrangement~\cite{PRB,A_G}. This, on turn, affects the characteristic relaxation time of the CRR through
\begin{equation}\label{1}
    \tau \sim e\,^{\Delta F/k_B T} \,,
\end{equation}
where $k_B$ is the Boltzmann constant and $T$ is the absolute temperature.

In a similar manner we shall consider small amorphous regions embedded by crystals. In general, however, $\Delta F$ does not only depend on the conditions at the boundary of the confining volume, but also on the actual chain conformation within the CRR.
In other words, once a conformational transition has occurred, the free energy barrier in the new state in general differs from its former value. This already suggests a possible origin of the connection between the mean barrier $\Delta F$ and its dispersion $\delta (\Delta F)$ in an ensemble of CRR's.

The description of the evolution dynamics of $\Delta F$ in a CRR is a difficult task. For this reason we shall limit ourselves to propose a statistical mechanical model with the aim of pointing out the characteristic features that an ensemble of CRR's in a semicrystalline polymer should have in order to meet the general principles of thermodynamics. This will lead us to the derivation of a distribution of constraining conditions, i.e., of single time decay processes contributing to the overall segmental relaxation process.

As an example of application, this distribution will be used to analyze some dielectric relaxation data of PET already
available in the literature.

\section{Phenomenological model}

\subsection{Thermodynamic considerations}

A finite amorphous domain in a polymer is a thermodynamic subsystem which, from the statistical mechanical point of view, can be thought to be made of two parts; one consists of only the chain conformational degrees of freedom (C-component), while the other (V-component) collects the vibrational ones. The observation that the specific heat of a glass is very close to that of the corresponding crystal (wherever the system is able to crystallize) suggests that these two components are approximately independent. In other words,
\begin{equation}\label{2}
	Z \simeq Z_{vib}\,Z_{conf} \,,
\end{equation}
where $Z$, $Z_{vib}$ and $Z_{conf}$ are the partition functions of the whole subsystem, the vibrational component and the conformational one respectively. However, there are processes where the direct C-V interaction is important and approximations like eq.~\ref{2} must be handled with care. Segmental relaxation in confined regimes is one of these cases.
%============================================================================================================
\begin{figure}
\includegraphics[width=8.5 cm, angle=0]{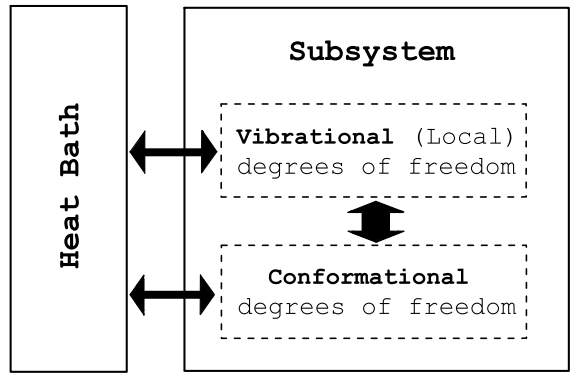}
\caption{\small{
Schematic of an amorphous domain viewed as a thermodynamic subsystem. In the presence of confinement by crystals, the direct interaction of the conformational degrees of freedom with the heat bath vanishes.
}} \label{fig1}
\end{figure}
%============================================================================================================

The scheme reported in Fig.~1 represents our subsystem, with its two components C and V, interacting with the heat bath (HB). The V component of the subsystem always interacts directly with both the HB and the conformational part C. On the other hand, the direct coupling of the latter with the HB is strongly affected by confinement. When the subsystem is embedded by crystals, for instance, the conformational degrees of freedom are isolated from the HB, the conformational fluctuation modes with wavelength larger than the characteristic dimensions of the subsystem are depressed and the C-HB coupling is rendered possible only through the C-V interaction. Hence, the latter is responsible of the fluctuations occurring in the conformational component under crystal confinement.

Of course the above one is an idealized picture. In the case of polymers, vibrational and conformational degrees of freedom are not so sharply separated with regards to confinement effects. Rather, we have to think that there are motions that are sensitive to boundary conditions, others that are not, and a relatively wide region of the spectrum in between. We can safely say that the modes for which confinement is uneffective are comparatively faster than the others, but we cannot exclude that suitably fast segmental modes belong to the intermediate class. This depends of course on how effective is the confinement and on the degree of cooperativity characteristic of the system. We shall comment this point further below.

As illustrated above, the effect of the crystals in regime of confined relaxation is the quenching of the conformational modes which are associated to density fluctuations with wavelengths larger than a certain value. In this respect, it is worth mentioning that there exists another situation where the long wavelenght fluctuation modes appear depressed, namely, when their characteristic evolution time is so slow to appear frozen with respect to the observation time scale (this is in fact the mechanism involved in the glass transition). When the attempt is made of describing the apparent relaxation dynamics in terms of statistical thermodynamics, we are necessarily lead to consider the existence of virtual conformational constraints. The difference with respect to confinement by crystal consists in the absence of a localized volume, with ideal rigid walls, which could cause the quenching of the modes. In other words, the constraints are spread out over the entire liquid, as they are only of a kinetic nature. For this reason we call this mechanism \emph{kinetic} confinement as opposed to the former which we shall call \emph{static}.

In the case of kinetic confinement the direct C-HB coupling is depressed by the quenching of some interaction channels (i.e. the long wavelength modes), but does not totally disappear. In this situation the direct C-HB and V-C interactions could be of comparable relevance in affecting the relaxation dynamics.

In the following we shall consider cooperatively rearranging regions whose confinement is mainly of a static nature, in the sense that wherever the border of these regions is not coincident with the surface of a crystal, the heat transfer through this "hole" is negligible with respect to the transfer via the V-C coupling. For convenience we shall call any of them static-CRR (SCRR). The case of the kinetic confinement regime is left to future work.

\subsection{The mean field distribution of constraints}

%Let $Z_{u}$ be the partition function associated to the conformational fluctuations in a (constant) volume $V$ within an unconstrained amorphous region of a polymer. The presence of constraints determines the emergence of a fluctuation energy threshold $E_{min}(V)$ below which conformational rearrangements are not possible. The reduced partition function $Z_c (E>E_{min})$, obtained by considering only the fluctuations allowing rearrangement, is smaller than $Z_{u}$;

Consider a (small) domain of volume $V$ in an unconstrained amorphous region of a polymer, and let $Z_u$ be the partition function associated to it. The emergence of conformational constraints in this domain has two main consequences. On the one hand, the partition function decreases to $Z_u '< Z_u$ because some states are ruled out; on the other, the transition from one conformational state to another in general involves the passage through non-equilibrium states, which means that a fluctuation free energy barrier $\Delta F \geq 0$ has to be overcome in order that the conformational change takes place. From the statistical mechanical point of view, the set of these non-equilibrium states is described by a reduced partition function, $Z_c$, collecting only the states (among all those contributing to $Z_u '$) which allow for conformational rearrangements~\cite{A_G}:
\begin{equation}\label{3}
	\Delta F = -k_B T \ln\frac{Z_c}{Z_{u}'} \,.
\end{equation}

The deviation from an initial (metastable) equilibrium state of a SCRR - a subsystem in contact with a heat bath - is in general accompanied by an increase of its energy (see e.g. ref.~\cite{Landau}, Sect.~21). Within the spirit of a mean-field description of the conformational transition process, we follow the suggestion indicated by ref.~\cite{A_G} and take $Z_c$ as the part of $Z_u '$ consisting of just the terms with energy $E \geq E_{min}$, being $E_{min}$ a threshold value below which conformational rearrangements are not possible. Note that the establishment of an $E_{min}$ is related to the transition process only. Of course, there may be cases where the actual $E_{min}$ and the actual energy value $E_i$ of the initial conformational state coincide, but on average the relation $E_{min}>E_i$ holds.
In our mean field treatment the energy threshold $E_{min}$ will be taken as a measure of the level of constraints.

We now consider an ensemble of SCRR's with a certain distribution of shapes and dimensions, that is, the number of monomers within the SCRR's is distributed around an effective mean $z$. For convenience, in the following, we shall refer the extensive quantities to the monomers - as the smallest crystallizable units - and deal with chemical potentials instead of free energies.

In connection with the cooperativity character of conformational rearrangement, all monomers in a SCRR are equivalent, so we can associate to each of them a specific partition function $Z\equiv [Z_{SCRR}]^{1/z}$, where $Z_{SCRR}\equiv Z_{SCRR}(z)$ is the partition function associated to the whole SCRR. The specific partition function $Z$ collects the complexity of the monomer viewed as a thermodynamic subsystem; furthermore, it implicitly includes the interaction with the other monomers forming the SCRR. Associating to each monomer a specific partition function, has the same meaning as associating to it a chemical potential (cf. ref.~\cite{A_G}).

In general, when a conformational change occurs in a SCRR, the new state differs from the parent one also in the value of the lower energy bound per monomer, $\zeta$, separating the fluctuations that can induce rearrangement from the others which cannot. With regards to the (instantaneous) characteristic relaxation time $\tau_{SCRR}$, the energy $\zeta$ suitably defines the state of a SCRR or, which is the same, of all the monomers of which is made; the reason is that different chain conformations yield the same $\tau_{SCRR}$ provided the associated $\zeta$ values are the same.

We now distribute all monomers into different classes, labelled by the value of $\zeta$ characterizing the SCRR's to which they belong. At this point the individuality of the SCRR's is lost and we remain with the problem of finding the most probable monomers' distribution $p(\zeta)$.

It is clear that any change in the distribution is rendered possible by the monomers which undergo a fluctuation to a state with an energy larger than the threshold $\zeta$ labelling the class to which they belong; let's call "active" these monomers. Once this threshold is exceeded, a monomer may change its state and fall into a different class; this is to say that the SCRR to which it belongs has reached a different conformational state. We cannot say which class an active monomer will fall in after each transition (it can even remain in the same class), but in a stationary state the ensemble of all active monomers will be distributed among the different classes in the most probable way.

In order to find this distribution, we consider the whole set of active monomers as a thermodynamic subsystem in thermal equilibrium with the heat bath. The total volume of this subsystem is constant (it is related to the overall crystallinity), so the most probable distribution can be found by minimizing the associated Helmholtz potential
%To this aim, we follow the classical procedure of extremizing the relevant thermodynamic potential under appropriate conditions. Since the volume of each SCRR and the total number of monomers in the amorphous state are constant, we need to minimize the Helmholtz potential
\begin{equation}\label{4}
	A = U - TS \,,
\end{equation}
where $S$ is the entropy and $U$ is the average energy involved in the transfer of monomers from one class to the other.
More explicitly, we associate to each class the average energy $\left\langle E\right\rangle_\zeta$ of the corresponding active monomers; then, $U$ is a suitable average of these terms (see below).

On one hand, the entropy term takes the customary form for a distribution of units in the different classes
\begin{equation}\label{5}
	S = - k_B \sum_{\zeta} p(\zeta) \ln p(\zeta) \,;
\end{equation}
on the other, the average energy involved in the transfer of units between classes is
\begin{equation}\label{6}
	U = \sum_{\zeta} p(\zeta)w_{r}(\zeta)\left\langle E\right\rangle_{\zeta} \,,
\end{equation}
where the average energy $\left\langle E\right\rangle_\zeta \,$, associated to the monomers able to leave their class ($\zeta$), is related to the reduced specific partition function
\begin{equation}\label{9}
	Z_{\zeta}\equiv \sum_{m;\,E_m\geq \zeta}e^{-E_m/k_B T}
\end{equation}
(being $E_m$ the energy of the $m$-th state) by
\begin{equation}\label{7}
	\left\langle E\right\rangle_\zeta = - \frac{\partial}{\partial (k_B T)^{-1}} \ln Z_{\zeta} \,;
\end{equation}
the factor
\begin{equation}\label{8}
	w_{r}\equiv \frac{Z_{\zeta}}{Z_0} \,,
\end{equation}
with $Z_0 \equiv Z_{\zeta}(\zeta=0)$, is the probability that a fluctuation increases the monomer energy above the actual minimum threshold for a rearrangement.

The equilibrium distribution minimizes the Helmholtz potential eq.~\ref{4} under the condition
\begin{equation}\label{10}
	\overline{\Delta\mu} \equiv \sum_{\zeta} p(\zeta)\,\Delta\mu(\zeta) = const. \,,
\end{equation}
where the chemical potential barrier is given by
\begin{equation}\label{11}
	\Delta\mu(\zeta) = -k_B T \ln\left( \frac{Z_{\zeta}}{Z_0}\right) \,.
\end{equation}
This condition simply says that, independent of how the exchange of elements among classes occurs, the thermodynamic state of the liquid, i.e. the average level of conformational constraints (to which the rearrangement chemical potential barrier height $\overline{\Delta\mu}$ is associated) is stationary, at least within the time scale of the observation. Shouldn't this condition being fulfilled, we would envisage the possibile variation of the crystallinity during the observation.

Straightforward calculations yield the following form for the most probable distribution:
\begin{equation}\label{12}
	p(\zeta) = N(\lambda) e^{-w_{r}(\zeta)\left\langle E\right\rangle_{\zeta}/k_B T - \lambda\Delta\mu(\zeta)} \,,
\end{equation}
where $N(\lambda)$ is the normalization constant and $\lambda$, the Lagrange multiplier associated to the condition eq.~\ref{10}, is related to the average chemical potential barrier $\overline{\Delta\mu}$ by the implicit relation
\begin{equation}\label{13}
	\overline{\Delta\mu}=-\frac{\partial}{\partial\lambda} \ln\left[\sum_{\zeta \geq 0}e^{-w_{r}(\zeta)\left\langle E\right\rangle_{\zeta}/k_B T - \lambda\Delta\mu}\right] \,.
\end{equation}

The fact that the probability factor $w_{r}$ is not contained in the entropy term may appear inconsistent; however, this is not the case. The point is that any monomer whatsoever, independent of whether it will change its state or not, is an element of a class; for this reason the entropy must be written in the above form eq.~\ref{5}. On the other hand, not all the elements of a class undergo a change of state to another class, but only the fraction $w_{r}(\zeta)$. To see this more clearly, consider that the equilibrium distribution could be as well obtained by extremizing $S$ with the condition on $\overline{\Delta\mu}$, eq.~\ref{10}, together with the conditions of constancy (stationarity) for the average energy involved in the exchange of elements among the classes, i.e. $U = const.$ The expression of the average energy, of course, has to take into account that not all the elements of a class [$\propto p(\zeta)$] undergo a transition, but only the fraction $w_{r}(\zeta)$ of them.

It is easy to show that also in the framework of this thermodynamics of constraints, the variance of $\Delta\mu$ can be expressed in terms of the mean value $\overline{\Delta\mu}$ by means of the expression
\begin{equation}\label{14}
	[\delta(\Delta\mu)]^2 \equiv\left\langle \Delta\mu^2 \right\rangle - \overline{\Delta\mu}^2 = -\frac{\partial\overline{\Delta\mu}}{\partial\lambda} \,.
\end{equation}
This equation is important, because it is related with the observations reported in the introductory section concerning the connection between free energy barrier height and heterogeneity, both introduced by the constraints.
When the $\lambda\Delta\mu$ term dominates in the exponent of eq.~\ref{12}, the distribution approaches a poissonian form. This is to say that the square root of the variance and the mean $\Delta\mu$ value tend to be the same in this limit.

\subsection{Relaxation function}

Segmental motion underlies the fluctuation regression of quantities like, for example, density or electric polarization. A key quantity for the description of these processes is the relaxation function $\phi (t)\equiv \left\langle B(t)B(0) \right\rangle/[B(0)]^2$, that is, the normalized autocorrelation function of a time dependent quantity $B(t)$ with respect to its value at the initial time $t=0$.

In general, cooperative rearrangement characterizes the segmental relaxation dynamics at long times~\cite{Sillescu}. However, in the case of \emph{static} confinement, where the kinetic component in the conformational constraints is small, the cooperative character of the motion is expected to extend down to comparatively shorter times. Considering the confined segmental motion only, we represent the relaxation function as a superposition of single-time decay contributions:
\begin{equation}\label{15}
	\phi(t)=\int d\zeta \, p(\zeta) \exp \left\{-\frac{t}{\tau(\zeta)}\right\} \,,
\end{equation}
where the characteristic relaxation time,
\begin{equation}\label{16}
	\tau(\zeta) = \tau^* e^{\,\Delta F(\zeta)/ k_B T } \,,
\end{equation}
carries information on cooperativity through the free energy barrier
\begin{equation}\label{16a}
	\Delta F(\zeta) = z\,\Delta \mu (\zeta)
\end{equation}
associated to the rearrangements of the SCRR's, and $\tau^*$ is a phenomenological triggering time which is expected to be rather smaller than the central relaxation time $\tau_0$ of the $\alpha$-process. Indeed, we showed at the beginning of this section that, if static confinement is effective, the conformational transitions must be stimulated by much faster (local) process.

The above expression of $\phi(t)$ still remains undetermined unless $Z_{\zeta}$ is known. On the other hand, the calculation of this partition function is in general not trivial at all and requires further modeling (which is out of the scope of the present contribution).

Alternatively, with the aim to illustrate a practical application of the model, we express the partition function in the form
\begin{equation}\label{17}
	Z_{\zeta} = \int^{\infty}_{\zeta} dE \rho(E,n)\, e^{-E/k_B T}\,\equiv Z_{\zeta ,\,n}  \,
\end{equation}
where the density of states $\rho(E,n)$ accounts for degeneracy and amplitude of the energy intervals between successive levels and it is taken in the form
\begin{equation}\label{18}
	\rho(E,n) = \rho_0 E\,^n \,,
\end{equation}
with $n$ an integer to be determined by fitting, and $\rho_0$ an irrelevant constant (cf. eqs.~\ref{7}, \ref{8} and \ref{11}). In fact, $n$ selects the relevant energy interval of the states contributing to $Z_u '$. Note that the above density of states, eq.~\ref{18}, is the same as for a set of $n+1$ independent harmonic oscillators in contact with a heat bath~\cite{greiner}.

In the case that $n=0$ the free energy barrier $\Delta F$ takes the form
\begin{equation}\label{19}
	\Delta F(\zeta) \simeq z\, \zeta \,.
\end{equation}
We recall that this functional dependence has been adopted to treat the case of strong confinement (i.e. the present \emph{static} confinement) in ref.~\cite{JCP_PET} on the basis of qualitative considerations.
Equation~\ref{19} states that the free energy of cooperative rearrangement has no entropy contribution. While this may be reasonable at high $\zeta$ values, it is not an acceptable approximation when $\zeta$ is small, i.e., when the unconstrained liquid state is approached.

\section{Example of application}

In this section we present the analysis of dielectric relaxation data as an example of application of the above model. We shall consider the cases of PET cold-crystallized from either a nematic or an isotropic glass. The data have been already published and the details of the experimental procedures, together with some aspects of the data analysis, can be retrieved from the literature~\cite{Tiberio_PET,JCP_PET}. We only recall that, as a first step, the complex dielectric permittivity $\varepsilon = \varepsilon' -i\varepsilon''$ is measured as a function of the angular frequency $\omega$. Then, the data are fitted to an empirical expression where each relaxation process is described by a corresponding Havriliak-Negami term, i.e.
\begin{equation}\label{20}
	\varepsilon_j(\omega)=\frac{\Delta\varepsilon_j}{\left[1+(i\,\omega\,\tau_j)^{a_j}\right]^{b_j}} \,,
\end{equation}
in which $j$ labels the relaxation process, $\Delta\varepsilon_j$ is the strength of the relaxation, $\tau_j$ is the central relaxation time, $a_j\in\,]\,0,\,1]$ describes the width of the relaxation ($a_j=1$ for the narrowest) and $b_j\in\,]\,0,\,1]$ its asymmetry ($b_j=1$ for a symmetric relaxation). The relaxation function associated to the $j$-th process is then obtained from the corresponding Havriliak-Negami component by means of a Fourier cosine transform~\cite{Will}, i.e.
\begin{equation}\label{21}
		\phi_j(t) = \frac{2}{\pi}\int^{\infty}_{0}\frac{\varepsilon_j ''(\omega)}{\Delta\varepsilon_j}\cos (\omega t)\frac{d\omega}{\omega} \,.
\end{equation}

%============================================================================================================
\begin{figure}
\includegraphics[width=8.5 cm, angle=0]{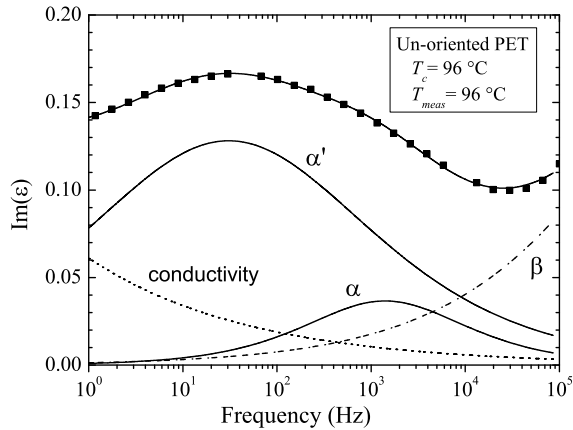}
\caption{\small{
Imaginary part of the dielectric permittivity at the temperature $T_{meas}=96$~$^{\rm{o}}$C, resolved in components,
of an unoriented PET sample cold-crystallized at $T_c=96$~$^{\rm{o}}$C for 360 min. The squares are the experimental values
}} \label{fig2}
\end{figure}
%============================================================================================================

In the following examples, we shall consider the $\alpha$-process emerging in the presence of a finite crystallinity. This relaxation has been named $\alpha '$ in ref.~\cite{Tiberio_PET} (it corresponds to the $\alpha_s$ process in ref.~\cite{JCP_PET}) and is believed to be a manifestation of static confinement. The faster $\alpha$-relaxation is active also in the absence of crystallinity and, wherever present in the crystallized system, is rather associated to kinetic confinement and will not be discussed here.

Figure~2 reports the imaginary part of the dielectric permittivity $\varepsilon ''$, as a function of the frequency $\nu = \omega/2\pi$, of an initially isotropic PET sample cold-crystallized for 360 min at a temperature $T_c = 96$~$^{\rm{o}}$C~\cite{Tiberio_PET}. The different components derived from the fitting procedure are also shown, namely, the two distinct $\alpha$ processes, the tail of the $\beta$-relaxation and the conductivity contribution. 

With respect to the $\alpha$ relaxation, the $\alpha '$ component is broader and centred at a lower frequency; this indicates that the latter process is influenced by constraining conditions to a larger extent than the former.

Figure 3 shows the $\varepsilon ''$ vs. $\nu$ dependence at a temperature $T=130$~$^{\rm{o}}$C in the case of a cold-drawn PET sample cold-crystallized at $T_c = 140$~$^{\rm{o}}$C for 120 min. Comparing it with the relaxation spectrum of the initially isotropic sample, we observe that the $\alpha$ process is absent and that the $\alpha '$ contribution is even broader than the corresponding one in Fig.~2. The central relaxation times of the two $\alpha '$ components reported in Figs.~2 and 3 are similar, that is, at the same $T$ the $\alpha '$ process in oriented PET is much slower than in initially isotropic PET.

%============================================================================================================
\begin{figure}
\includegraphics[width=8.5 cm, angle=0]{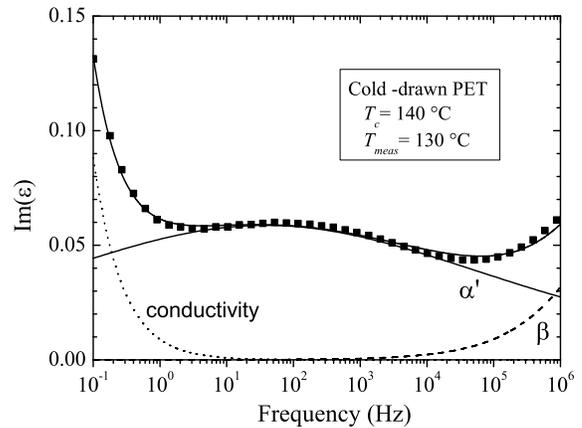}
\caption{\small{
Imaginary part of the dielectric permittivity at the temperature $T_{meas}=130$~$^{\rm{o}}$C, resolved in components,
of a cold-drawn PET sample cold-crystallized at $T_c=140$~$^{\rm{o}}$C for 120 min. The squares are the experimental values.
}} \label{fig3}
\end{figure}
%============================================================================================================

Figure 4 reports in a log-log plot the relaxation functions (open squares and triangles) associated to the $\alpha '$ processes of Figs.~2 and 3, obtained after eq.~\ref{21} (the inset shows the same in a lin-log plot). The solid lines are the best fitting results obtained by using the function defined by eq.~\ref{15}, which in a more explicit form reads:
\begin{eqnarray}
\nonumber
	\phi_{\alpha '}(t) = N(\lambda , n)\int_0^{\infty} d\zeta 
	\left(\frac{Z_{0,n}}{Z_{\zeta ,n}}\right)^{\lambda k_B T} \times  \\
\nonumber
	\exp \left\{
	-\frac{Z_{\zeta ,n+1}}{k_B T Z_{0 ,n}} - \frac{t}{\tau^*}\left(\frac{Z_{\zeta ,n}}{Z_{0,n}}\right)^z
	\right\}   \\
\nonumber
	\\
	\equiv\phi_{\alpha '}(\lambda, \tau^*, z, n; t)\,
\label{15fit}
\end{eqnarray}
where the fitting parameters have been explicitly indicated in last row, $Z_{\zeta ,n}$ is defined by eq.~\ref{17} above and the term corresponding to the average energy in the exponent of eq.~\ref{12} has been expressed as $w_{\zeta ,n}\left\langle E\right\rangle_{\zeta ,n}=Z_{\zeta ,n+1}/Z_{0 ,n}$; note that the normalization factor $N$ now depends on $n$ (cf. eq.~\ref{12}).

%============================================================================================================
\begin{figure}
\includegraphics[width=9 cm, angle=0]{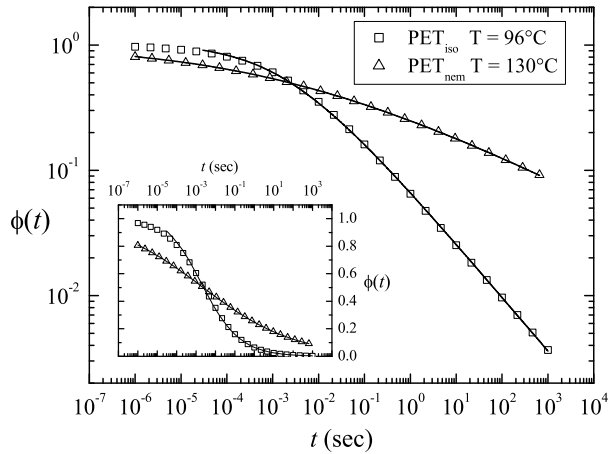}
\caption{\small{
Relaxation functions $\phi(t)$ associated to the static constrained $\alpha$-processes of PET reported in Figs.~2 and 3, as obtained from eq.~\ref{21}. The solid lines are the best fittings from eq.~\ref{15fit}. The inset shows the same using linear ordinates.
}} \label{fig4}
\end{figure}
%============================================================================================================

The analysis is carried out by first setting a time lower limit, say $t_{min}$, and fitting only the data for $t\geq t_{min}$; $n$ is assigned also a starting value, say $n=0$, which is set fixed during each fitting run. Then, the other parameters ($\lambda, \tau^*$ and $z$) are let free to adjust. The procedure then consists in lowering $t_{min}$ progressively (while trying different integers $n$) so to reach the minimum possible value compatible with an acceptable $\chi ^2$ (of course, the condition $\tau^* < t_{min}$ must always hold). The reliability of the results can be assessed by checking if the approximate equality (cf. eq.~\ref{16})
\begin{equation}\label{22}
	k_B T \ln\left(\frac{\tau_0}{\tau^*}\right) \approx z \overline{\Delta\mu}
\end{equation}
is verified by the final values of the fitting parameters.

We report in Table~I the values of crystallization temperature $T_c$, measurement temperature $T_{meas}$, width parameter $a$ (eq.~\ref{20}) of the relaxation (in all cases $b=1$) and central relaxation time $\tau_0$, together with the fitting parameters $\lambda$, $\tau^*$, $z$ and $n$ worked out by fitting $\phi(t)$ for $t\geq t_{min}$; in the same table are also reported $t_{min}$, the average energy threshold $\overline{\zeta}$, the dispersion $\delta\zeta$ around it, the average chemical potential $\overline{\Delta\mu}$ and the dispersion around it, $\delta[\Delta\mu]$ (last quantities being calculated as ordinary averages using the distribution $p(\zeta)$ once $\lambda$ and $n$ have been found).

The values reported in the Table verify eq.~\ref{22} above. Note that a more effective cooperativity (larger $z$) leads to smaller $\tau^*$ values at fixed $\tau_0$ and $\overline{\Delta\mu}$ (see Table~I).

There are a few points that deserve consideration. First of all, we always find that $\tau^* \ll \tau_0$; this is consistent with the idea that the observed $\alpha '$ relaxation is indeed stimulated by faster processes. Furthermore, we can see that in the case of oriented PET the number of monomers of a SCRR is significantly larger than in the case of initially isotropic PET, although $\zeta$ has a lower value (cf. Table~I). This is an important result, because it is consistent with the observation that the vitrification of amorphous domains induced by (cold-) crystallization is indeed rather more pronounced in oriented than in isotropic PET~\cite{PRB}. Cold-drawing was indeed shown to affect the density autocorrelation length of PET, $\xi$, in the sense of increasing its value along the orientation direction, i.e. $\xi_{nem}\simeq 2\,\xi_{iso}$~\cite{PRB}. The present results thus suggest a very clear connection between cooperativity and $\xi$.
The fact that $\xi$ is a decreasing function of the temperature~\cite{Edwards,PRB} further supports this results, since $z_{iso}|_{T=96^{\rm{o}}\text{C} }<z_{nem}|_{T=130^{\rm{o}}\text{C} }$.

About the actual $z$ values worked out by fitting, two aspects have to be taken into account. The first is that they come out from a mean field model, which means that they are not expected to be exact. The second is that they represent an underestimate when a (weak) direct coupling between the SCRR conformational degrees of freedom and the heat bath is effective: faster fluctuation decays are attributed to smaller SCRR's, within the model, because relaxation mechanisms other than through the local degrees of freedom are ruled out.

We conclude this section by comparing the present results with those of our previous analysis on oriented PET~\cite{JCP_PET}. In that article we guessed that a gaussian distribution of constraining conditions could be a reasonable description of time heterogeneity; moreover, we approximated the barrier free energy $\Delta F$ like in eq.~\ref{19} above. The resulting fitting function was rather simpler than that used here for the analysis of the data. However, the direct derivation of $z$ and $\Delta F$ from the fitting procedure was not possible, although the latter quantity could be estimated from $\tau^*$ and $\tau_0$ assuming that a relationship like eq.~\ref{22} would hold. We thus found that $k_B T \ln(\tau_0/\tau^*) \simeq 11.9$ kcal/mol~\cite{JCP_PET}, to be compared with the present value of 8.9 kcal/mol.

On the other hand, the estimate of the free energy barrier dispersion from the fitting of the relaxation function, as reported in ref.~\cite{JCP_PET}, yielded $\delta (\Delta F) \simeq 8.8$ kcal/mol, while in the present case we obtain $z \delta[\Delta\mu] \simeq 6.7$ kcal/mol. This is to say that the former approach, ref.~\cite{JCP_PET}, can be safely used for an estimate of $\delta (\Delta F)$ in the static confinement regime; more noticeably, already that simple analysis indicated that $\delta (\Delta F)$ and $\Delta F$ were comparable, a circumstance that has been verified to occur also in other semicrystalline polymers~\cite{PPA_PPS}.

\section{Concluding remarks}

A central result of this paper is the connection between the average free energy barrier $\Delta F = z\overline{\Delta\mu}$ associated to the conformational rearrangement within the SCRR's and the width $\delta F\equiv z\,\delta(\Delta\mu)$ of the barrier height distribution in an ensemble of SCRR's:
\begin{equation}\label{25}
	\Delta F \sim \delta (\Delta F) \,.
\end{equation}
In the introductory section, it was already pointed out that the increase of the barrier height in conformational rearrangement is accompanied by an increase in the heterogeneity of constraining conditions. Furthermore, from an analysis of the relaxation behavior of PET in ref.~\cite{JCP_PET} it was found that a relationship like eq.~\ref{25} could hold.
The theoretical model developed above represents an attempt to interpret this phenomenology.

The classical theory of thermodynamic fluctuations (cf. e.g. ref.~\cite{Landau}) already establishes a connection between the average value of an extensive quantity and the amplitude of its fluctuations: as the volume of a subsystem decreases, the latter approaches the former. Analogously, the application of the present model to the analysis of the relaxation function, indicates that a relationship like eq.~\ref{25} characterizes the fluctuations of constraints on subsystems consisting of $z$ monomers, which in fact turn out to be small.

In the present model, the heterogeneity of the SCRR ensemble arises as a consequence of thermodynamic equilibrium wherever static confinement is effective. Of course, the question
whether conformational relaxation under confinement (either static or dynamic) consists of a superposition of single time decay processes, or it is the manifestation of an intrinsically non exponential decay, is not answered presently. This is just because of the thermodynamic nature of the model: a framework where the dynamics of monomer transfers among different $\zeta$-classes is disregarded.

We now turn to an aspect that we pointed out already in the thermodynamics subsection, namely, the separation between relaxation modes that are sensitive or non-sensitive to the confining conditions. Comparing the analyses carried out above on the two PET samples, we observe that, although the central relaxation times $\tau_0$ are practically the same (i.e. 5 and 4.2$\times 10^{-3}$ sec), the fitting triggering times $\tau^*$ are significantly different; indeed, oriented PET is characterized by an enhanced cooperativity and a larger readjustment free energy barrier. On the other hand, the meaning of eqs.~\ref{15} and~\ref{16} is that the $\alpha '$ relaxation is the long time scale manifestation of fast processes through the coupling of the local degrees of freedom with the conformational ones. In a sense, there is no primary process defining for instance the central relaxation time $\tau_0$, because the latter just characterizes an asymptotic dynamics. Through the calculation of the free energy $\Delta F$, our analysis ultimately indicates how faster than $\tau_0$ can be a process for it being still affected by confinement (cf. eq.~\ref{22}). Below this lower bound ($\tau^*$) there may well be, depending on the cooperativity of the system, fast segmental modes which are local enough not to be affected by the presence of the crystals.

A number of open points still remain to go into thoroughly after this preliminar description of the restricted conformational dynamics, but probably the most important one would be the extension of the model to the case of the ordinary $\alpha$-process, where confinement has a kinetic character. We hope, however, that the present contribution may be a useful starting point for further progress in this field.

\section{Acknowledgements}

M.P. gratefully acknowledges MCYT for the award of a sabbatical grant (SAB-2006-0077) to work at IEM-CSIC, Madrid and CNR, Italy, for supporting this initiative within the framework "`Congedi per Motivi di Studio"'. T.A.E. is indebited to MCYT for grant MAT2005-01768

\clearpage

%==========================================================================================
\begin{table}%[h]
    \begin{center}
        \begin{tabular}{cccccccccccccc}
        \hline
        Sample & $T_c$ & $T_{meas}$ & $a$ & $\tau_0$ & $t_{min}$ & $\lambda$ & $\tau^*$ & $z$ & $n$ & $\overline{\zeta}$ & $\delta\zeta$ & $\overline{\Delta\mu}$ & $\delta[\Delta\mu]$  \\
               & ($^{\rm{o}}$C) & ($^{\rm{o}}$C) &  & (sec) & (sec) & (kcal/mol)$^{-1}$ & (sec) &  &  & (kcal/mol) & (kcal/mol) & (kcal/mol) & (kcal/mol)    \\
        \hline
        
         isotropic  & 96  & 96  & 0.4  & $5\times 10^{-3}$ & $3\times 10^{-5}$ & 1.45 & $2\times 10^{-5}$ & 2.5 & 5 & 6.5 & 2.1 & 1.65 & 1  \\

         nematic  & 140 & 130 & 0.18 & $4.2\times 10^{-3}$ & $10^{-6}$ & 1.92 & $5.4\times 10^{-8}$ & 7.6 & 5 & 5.7 & 2.3 & 1.23 & 0.88 \\
         
        \hline
        \end{tabular}
    \end{center}
    \caption{
Crystallization temperature $T_c$, measurement temperature $T_{meas}$, width parameter $a$, central relaxation time $\tau_0$ (cf. refs.~\cite{Tiberio_PET,JCP_PET}), lower fitting time bound $t_{min}$ and relaxation function parameters ($\lambda$, $\tau^*$, $z$ and $n$) of the $\alpha '$-processes in static confinement regime for PET samples crystallized from isotropic and nematic glasses; the average minimum energy threshold $\overline{\zeta}$, its dispersion around the mean $\delta\zeta$, the average chemical potential $\overline{\Delta\mu}$ and its dispersion $\delta[\Delta\mu]$ as calculated from the fitting parameters using the distribution eq.~\ref{12} are also reported.
     }
    \label{tab1}
\end{table}

\end{document}